\documentclass{amsart}

\usepackage[dvips]{graphicx}
\usepackage{amssymb}

\begin{document}

\title[Bit propagation in Josephson junction arrays]{On the transmission of binary bits in discrete Josephson-junction arrays}

\author{J. E. Mac\'{\i}as-D\'{\i}az}
\address{Departamento de Matem\'{a}ticas y F\'{\i}sica, Universidad Aut\'{o}noma de Aguascalientes, Aguascalientes, Ags. 20131, Mexico}
\email{jemacias@correo.uaa.mx}

\author{A. Puri}
\address{Department of Physics, University of New Orleans, New Orleans,  LA 70148}
\email{apuri@uno.edu}

\subjclass[2010]{(PACS) 02.60.Lj; 63.20.Pw; 74.50.+r}
\keywords{discrete Josephson-junction arrays; nonlinear supratransmission and infratransmission; signal propagation; sine-Gordon equation; localized modes}

\begin{abstract}
In this work, we use supratransmission and infratransmission in the mathematical modeling of the propagation of digital signals in weakly damped, discrete Josephson-junction arrays, using energy-based detection criteria. Our results show an efficient and reliable transmission of binary information.
\end{abstract}

\maketitle

\section{Introduction}

The numerical study of wave transmission in long Josephson structures subject to harmonic driving was initiated  in the middle 1980's by Olsen and Samuelsen \cite{Olsen}, and the investigation was extended to coupled arrays of short superconducting tunnel junctions \cite {Barday}. The study of the nonlinear bistability in Josephson junctions was the central topic of research in these works, a study that was continued later on via perturbation analysis with partially satisfactory results \cite {Kivshar} until, finally, the whole analytical apparatus for the undamped, continuous-limit case was recently unveiled \cite{Chevrieux2}. Nonetheless, the investigation in the area has proved to be rather rich and interesting \cite{Olsen2,Kivshar2,Makhlin,Makhlin2,Makhlin3,Geniet-Leon}.

In this paper, we develop an application of nonlinear supratransmission and infratransmission to the propagation of binary information in weakly damped, discrete, finite Josephson-junction arrays, by modulating the amplitudes of the driving signal at one end. The first section of this letter introduces the mathematical model under study and the energy expressions associated with the problem, while the next section briefly presents the finite-difference schemes employed to approximate the model. A computational study to predict the occurrence of nonlinear supratransmission under the presence of external damping is carried out in the next section. Next, we study briefly some characteristics of the kinks emitted by the boundary, and the following section describes the technique employed in order to transmit binary information in discrete Josephson junction arrays, presenting a simulation of transmission. Finally, we provide a section of concluding remarks.

\section{Mathematical model}

Throughout this paper, we assume that $\gamma$ and $c$ are nonnegative real numbers. Likewise, we consider a system $( u _n ) _{n = 1} ^N$ of $N$ Josephson junctions coupled through superconducting wires, satisfying the discrete, boundary-initial-value problem
\begin{equation}
\begin{array}{c}
\displaystyle {\ddot {u} _n - c ^2 \Delta ^2 u _n + \gamma _n \dot {u} _n + \sin u _n} = \mu \ \ \ (1 \leq n \leq N), \\ 
		\begin{array}{rl}
        \begin{array}{l}
            {\rm subject\ to:} \qquad \\ \\
        \end{array}
        \left\{
        \begin{array}{ll}
            u _n (0) = 0, & 1 \leq n \leq N, \\
            \displaystyle {\frac {d u _n} {d t} (0) = 0}, & 1 \leq n \leq N, \\
            u _0 - u _1 = \displaystyle {\frac {\phi (t)} {c ^2}}, & t \in (0 , + \infty), \\
            u _{N + 1} - u _N = 0, & t \in (0 , + \infty),
        \end{array}\right.
    \end{array}
\end{array}\label{Eqn:DiscreteMain20}
\end{equation}
where $\gamma _n = \gamma$ for every $n < N$, and $\gamma _N = \gamma + 1 / R$. The number $R$ is called the \emph {output reading resistance} of the system, and it is related to the output current intensity $I$ through Ohm's law: $I = \dot {u} _N / R$. The number $c$ will be called the \emph {coupling coefficient}, and $\gamma$ evidently plays the role of an external damping coefficient. The constant $\mu$ will be called the \emph {Josephson current} of the system by abusing the standard terminology, and its inception in our model corresponds to a mere mathematical generalization of the model describing discrete arrays of Josephson junctions. Indeed, if $\mu$ is equal to zero, then $\phi (t)$ and $I$ represent the input intensity function and the output current intensity, respectively, normalized to the Josephson critical current.

Notationally, $\dot {u}$ and $\ddot {u}$ denote the first- and the second-order derivatives of $u$ with respect to time, respectively, and $\Delta _x u _n$ represents the spatial forward-difference $u _{n + 1} - u _n$, while $\Delta ^2 = \Delta _x \Delta _{\bar {x}}$, where the backward-difference $\Delta _{\bar {x}} = u _n - u _{n - 1}$ is employed. The function $\phi$ may be assumed to be continuously differentiable over $(0 , + \infty)$ in general. The particular case when $\phi$ is a sinusoidal function will be of special interest. More concretely, we will study the case when the driving is given by the function $\phi (t) = A \sin (\Omega t)$.

It is worth noticing that our mixed-value problem may be approximated via a continuous, Neumann boundary-value problem involving a perturbed sine-Gordon model for strong coupling. Also, notice that if $\gamma = \mu = 0$, then the Hamiltonian of the $n$th lattice site is given by 
\begin{equation}
H _n = \frac {1} {2} \left[ \dot {u} _n ^2 + c ^2 (u _{n + 1} - u _n) ^2 \right] + 1 - \cos u _n. \label{Eq:Hamil}
\end{equation}
After including the potential energy from the coupling between the first two sites, the total energy of the system becomes
\begin{equation}
E = \sum _{n = 1} ^N H _n + \frac {c ^2} {2} (u _1 - u _0) ^2.
\end{equation}
A simple integration of this formula over a finite interval of time gives the total energy injected into the system over such interval.

The limiting case consisting of the discrete Josephson-junction array with an infinite number of junctions is of particular interest. In this situation, it is easy to verify that the rate of change of the energy of the system with respect to time is provided by the formula
\begin{equation}
\frac {d E} {d t} = \phi (t) \dot {u} _0 - \gamma \sum _{n = 1} ^\infty (\dot {u} _n) ^2.
\end{equation}

\section{Numerical scheme}

\paragraph* {First finite-difference scheme.} We consider the system of differential equations (\ref {Eqn:DiscreteMain20}), and a regular partition $0 = t _0 < t _1 < \dots < t _M = T$ of the time interval $[0 , T]$ with time step equal to $\Delta t$. For each $k = 0 , 1 , \dots , M$, let us represent the approximate solution to our problem on the $n$th lattice site at time $t _k$ by $u ^k _n$, and let us convey that $\delta _t u _n ^k = u _n ^{k + 1} - u _n ^{k - 1}$, that $\delta ^2 _t u _n ^k = u _n ^{k + 1} - 2 u _n ^k + u _n ^{k - 1}$ and that $\delta ^2 _x u _n ^k = u _{n + 1} ^k - 2 u _n ^k + u _{n - 1} ^k$. In order to posses discrete energy expressions that consistently approximate their continuous counterparts, the continuous-time equations will be approximated by the discrete expressions
\begin{equation}
\displaystyle {\frac {\delta ^2 _t u _n ^k} {(\Delta t) ^2} - c ^2 \delta ^2 _x u _n ^k + \frac {\gamma _n} {2 \Delta t} \delta _t u _n ^k +\frac {V (u _n ^{k + 1}) - V (u _n ^{k - 1})} {u _n ^{k + 1} - u _n ^{k - 1}}} = 0, \label{Eq:FDS}
\end{equation}
for every $n = 1 , \dots , N$. Here, $V (u) = 1 - \cos u$ at any time.

Our finite-difference scheme is consistent with our mixed-value problem, conditionally stable, and possesses energy-invariant properties. More precisely, if the energy of the system at the $k$th time step is computed using the expression 
\begin{equation}
E _k = \sum _{n = 1} ^{N - 1} H _n ^k + \frac {c ^2} {2} (u _1 ^{k + 1} - u _0 ^{k + 1}) (u _1 ^k - u _0 ^k),
\end{equation}
where the individual energy of the $n$th lattice site is computed by
\begin{eqnarray}
H _n ^k & = & \frac {1} {2} \left( \frac {u _n ^{k + 1} - u _n ^k} {\Delta t} \right) ^2 + \frac {c ^2} {2} (u _{n + 1} ^{k + 1} - u _n ^{k + 1}) (u _{n + 1} ^k - u _n ^k)\\ \nonumber
 & & \quad + \frac {V (u _n ^{k + 1}) + V (u _n ^k)} {2}, \label{Eq:DHamil}
\end{eqnarray}
then the discrete rate of change of energy turns out to be a consistent approximation for the corresponding instantaneous rate of change, while the approximations of the individual and the total energies are consistent estimates of their continuous counterparts. Particularly, the following property of consistency in the energy domain holds for every $k = 1 , \dots , M$:
\begin{equation}
\frac {E _k  - E _{k - 1}} {\Delta t} = \phi (t _k) \frac {\delta _t u _0 ^k} {2 \Delta t} - \gamma \sum _{n = 1} ^{N - 1} \left( \frac {\delta _t u _n ^k} {2 \Delta t} \right) ^2.
\end{equation}

\paragraph* {Second finite-difference scheme.} It is important to observe that the energy functional $H _n$ defined in (\ref {Eq:Hamil}) is positive definite; however, there is no guarantee that the corresponding discrete approximation given by (\ref {Eq:DHamil}) is also positive definite. In order to save that shortcoming, we present now the finite-difference scheme
\begin{equation}
\frac {\delta ^2 _t u _n ^k} {(\Delta t) ^2} - \frac {c ^2} {2} \bar {\delta} \delta ^2 _x u _n ^k + \frac {\gamma} {2 \Delta t} \delta _t u _n ^k + \frac {V (u _n ^{k + 1}) - V (u _n ^{k - 1})} {u _n ^{k + 1} - u _n ^{k - 1}} = 0,
\end{equation}
where $\bar {\delta} u _n ^k = u _n ^{k + 1} + u _n ^{k - 1}$. The local energy at the $n$th node and time $t _k$ is approximated then using the discrete expression
\begin{equation}
\begin{array}{rcl}
H _n ^k & = & \displaystyle {\frac {1} {2} \left( \frac {u _n ^{k + 1} - u _n ^k} {\Delta t} \right) ^2 + \frac {c ^2} {8} \left[ \left( u _{n + 1} ^{k + 1} - u _n ^{k + 1} \right) ^2 + \left( u _{n - 1} ^{k + 1} - u _n ^{k + 1} \right) ^2 \right.} \\
 & & \quad \displaystyle {+ \left. \left( u _{n + 1} ^k - u _n ^k \right) ^2 + \left( u _{n - 1} ^k - u _n ^k \right) ^2 \right] + \frac {V (u _n ^{k + 1}) + V (u _n ^k)} {2}},
\end{array}
\end{equation}
which is always nonnegative. Moreover, the total energy of the system is estimated through
\begin{equation}
E _k = \sum _{n = 1} ^{N - 1} H _n ^k + \frac {c ^2} {4} \left[ \frac {\left( u _1 ^{k + 1} - u _0 ^{k + 1} \right) ^2 + \left( u _1 ^k - u _0 ^k \right) ^2} {2} \right].
\end{equation}
This new scheme is clearly consistent with respect to problem (\ref {Eqn:DiscreteMain20}) in both the solution and the energy domains. Moreover, the identity
\begin{equation}
\frac {E _k - E _{k - 1}} {\Delta t} = c ^2 \left( \frac {\bar {\delta} u _0 ^k - \bar {\delta} u _1 ^k} {2} \right) \frac {\delta _t u _0 ^k} {2 \Delta t} - \gamma \sum _{n = 1} ^{N - 1} \left( \frac {\delta _t u _n ^k} {2 \Delta t} \right) ^2
\end{equation}
holds for every $k = 1 , \dots , M$. 

\paragraph*{Computational remarks.} In order to simulate a discrete array of Josephson junctions of infinite length, we consider a finite system (\ref {Eqn:DiscreteMain20}) of relatively large length $N$, in which each $\gamma _n$ includes both the effect of external damping described above and a simulation of an absorbing boundary slowly increasing in magnitude on the last $N - N _0$ oscillators. Specifically, we let
\begin{equation}
\gamma _n = \kappa \left[ 1 + \tanh \left( \displaystyle {\frac {2 n - N _0 + N} {2 \sigma}} \right) \right] + \gamma + \frac {1} {R} \delta _N (n),
\end{equation}
where $\delta _N (n)$ is equal to $1$ if $n = N$, and zero otherwise. In practice, we will let $\kappa = 0.5$, $\sigma = 3$, $N = 60$, and $N _0 = 50$.

Also, in order to avoid the generation of shock waves in a system like (\ref {Eqn:DiscreteMain20}) subject to harmonic driving $\phi (t) = A \sin (\Omega t)$, we opt for increasing slowly and linearly the driving amplitude from $0$ to its actual value $A$, during a fixed period of time about the initial instant. After that, the driving amplitude will assume the constant value $A$.

Both numerical methods have been employed to produce the results in this article. Our results have shown that both methods produce results that are in excellent agreement for small values of $\Delta t$, an observation that leads to conclude that the results in the following sections are intrinsic to the problem under study and not scheme-dependent.

Computationally, the implementation of these finite-difference schemes is accomplished via Newton's method, which in turn leads to implement Crout's technique with pivoting to solve tridiagonal linear systems. The stopping criterion employed in Newton's method was reached when the relative error of subsequent approximations was less than $1 \times 10 ^{- 6}$, and the norm employed was the infinite norm in the space $\mathbb {R} ^N$.

\section{Analysis of supratransmission\label{Sec3}}

The process of nonlinear supratransmission consists of a sudden increase in the amplitude of wave signals transmitted into a nonlinear chain by a harmonic disturbance at the end, irradiating at a frequency in the forbidden band gap. The existence of a nonlinear supratransmission threshold of the energy administered into a finite array of Josephson junctions satisfying (\ref {Eqn:DiscreteMain20}) for a harmonic function $\phi$ and driving frequency in the forbidden band gap region $\Omega < 1$, has been established in the continuous-limit case \cite{Chevrieux2} and numerically predicted for a discrete system. Using our numerical scheme, a prediction of the occurrence of nonlinear supratransmission can be approximated for every driving frequency in the forbidden band gap of the system of equations (\ref {Eqn:DiscreteMain20}), by estimating the value of the driving amplitude at which a drastic increase in the total energy of the system is detected. In the continuous-limit case, the driving amplitude $A _s$ at which supratransmission first starts is related to $\Omega$ through the relation 
\begin{equation}
A _s = 2 c (1 - \Omega ^2). \label{Eqn:PredictedBif}
\end{equation}

\begin{figure}[tbc]
\centerline{\includegraphics[width=0.9\textwidth]{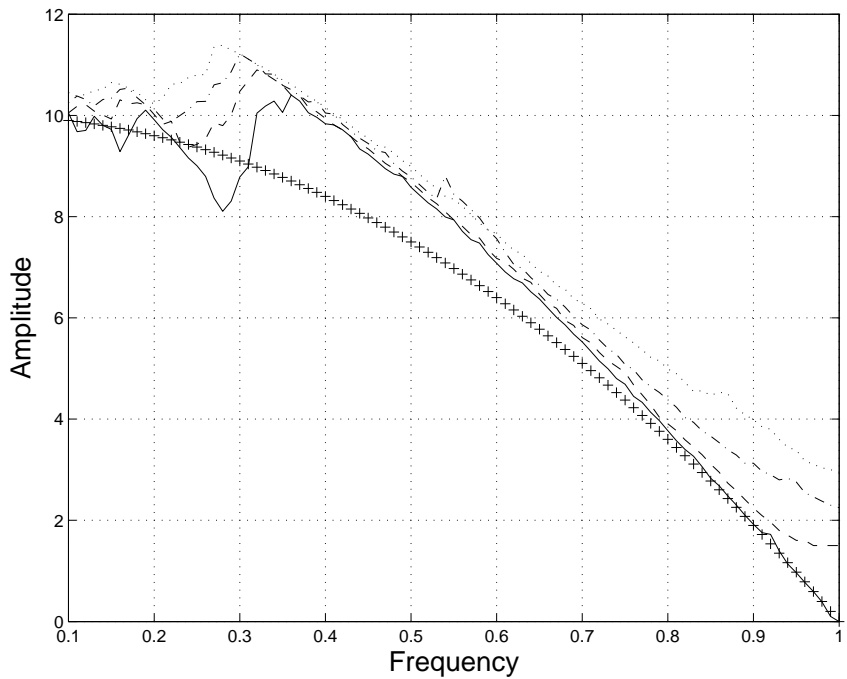}}
\caption{Bifurcation diagram of occurrence of critical amplitude versus driving frequency for problem (\ref {Eqn:DiscreteMain20}) with $c = 5$, for $\gamma = 0$ (solid), $0.1$ (dashed), $0.2$ (dash-dotted), $0.3$ (dotted). The continuous-limit prediction (\ref {Eqn:PredictedBif}) of the critical threshold is represented by a sequence of plus signs.\label{Fig:BifurcationGamma}}
\end{figure}

We follow next the standard methodology employed in \cite {Macias-Supra}, and consider several values of the damping coefficient. The results of our numerical computations are summarized as graphs of minimal amplitude at which supratransmission starts versus driving frequency, and they are presented in Figure \ref {Fig:BifurcationGamma}. The diagram representing the undamped scenario exhibits a good agreement with the continuous prediction for high frequencies, and damping is clearly shown to delay the appearance of the supratransmission threshold for frequencies $\Omega > 0.35$.

On the other hand, the process of nonlinear infratransmission or lower transmission, as opposed to supratransmission, consists in a sudden decrease in the amplitude of wave signals in a chain harmonically driven at its end. The existence of a nonlinear infratransmission threshold under which there is a drastic decrease in the energy injected into the system was established in \cite{Chevrieux2} in the continuous-limit case. As a matter of fact, the infratransmission threshold $A _i$ is less than the supratransmission threshold $A _s$, and between the amplitudes $A _i$ and $A _s$ lies a region of bistability. In order to approximate numerically the value of $A _i$ for a given driving frequency in the forbidden band gap with associated supratransmission threshold $A _s$, we chose a driving function $\phi (t) = A (t) \sin (\Omega t)$, with amplitude function defined via
\begin{equation}
A (t) = A _s (1 - e ^{- t / \tau _1}) + (A _0 - A _s) (1 - e ^{- t / \tau _2}),\label{Eqn:Amplitude}
\end{equation}
where $\tau _2 \gg \tau _1$. 

\begin{figure}[tbc]
\centerline{\includegraphics[width=0.9\textwidth]{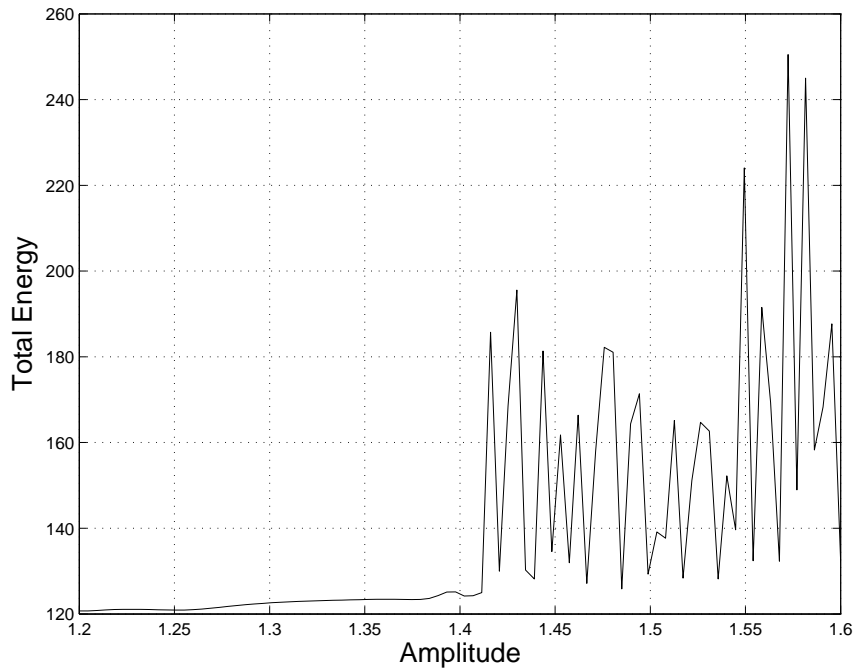}}
\caption{Graph of total energy vs. driving amplitude in an undamped, semi-infinite, discrete Josephson-junction array, driven at a frequency of $\Omega = 0.9$ via expression (\ref {Eqn:Amplitude}) during a period of time equal to $6000$, with a coupling coefficient of $5$.\label{Fig:InfraGraph}}
\end{figure}

\begin{figure}[tbc]
\begin{tabular}{ccc}
\includegraphics[width=0.45\textwidth]{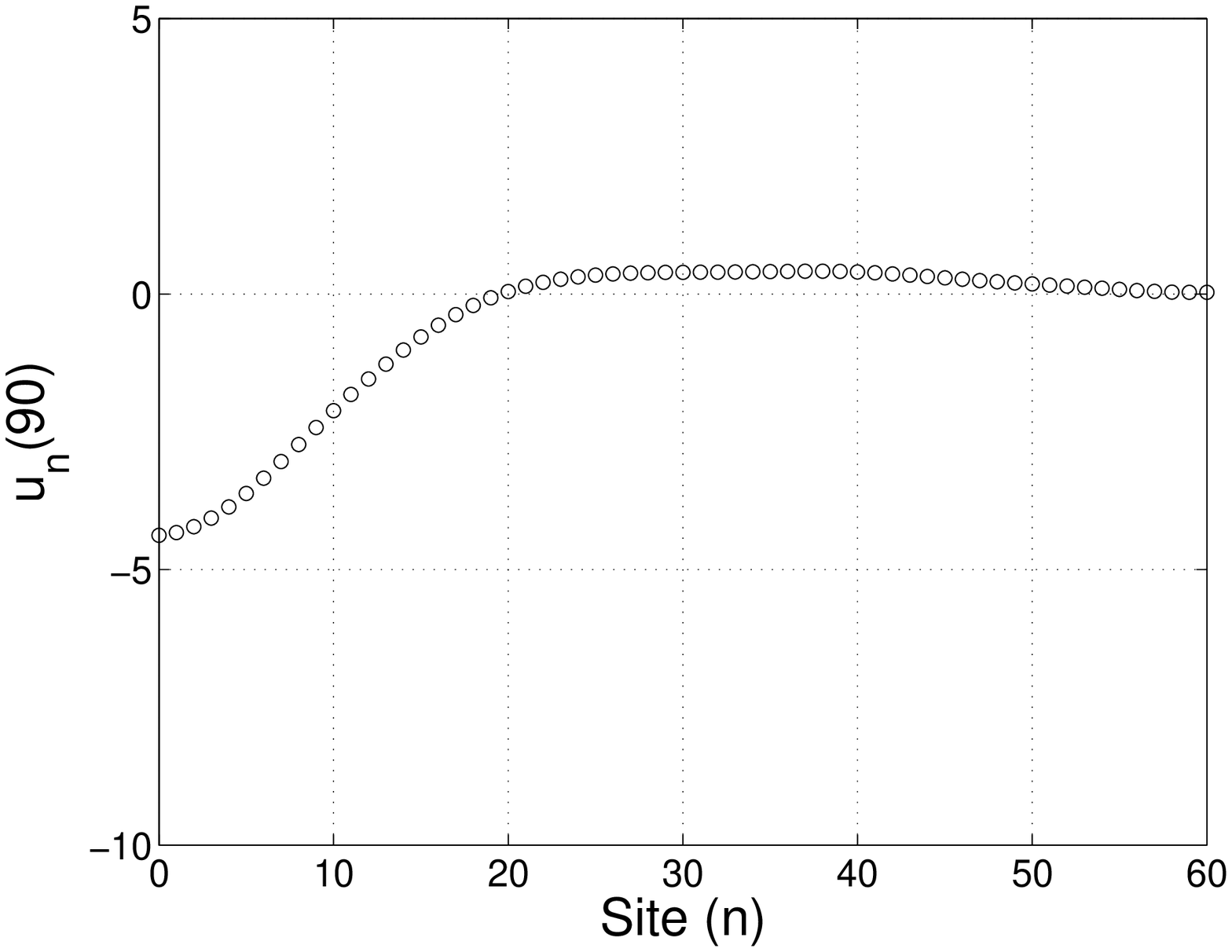}&\includegraphics[width=0.45\textwidth]{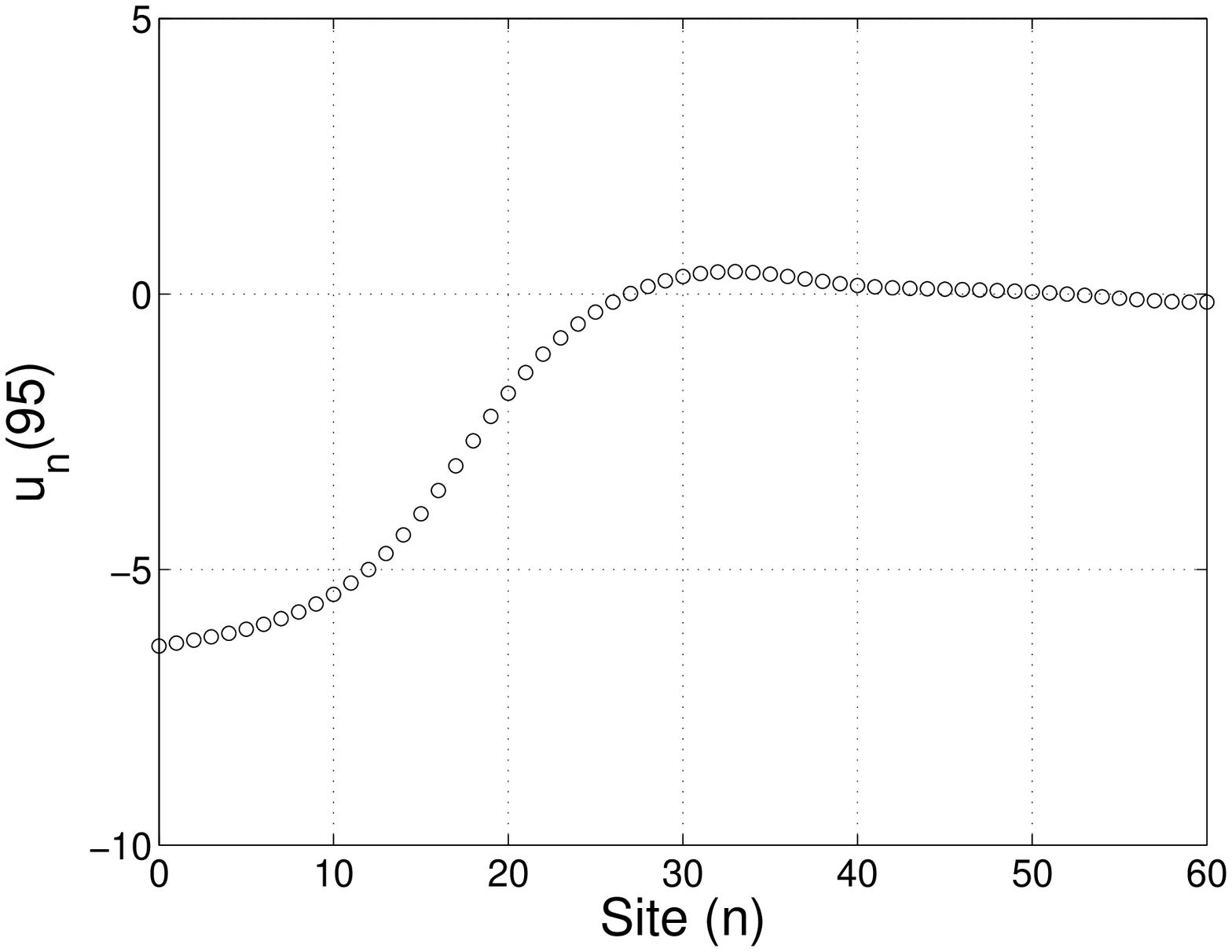}\\
\includegraphics[width=0.45\textwidth]{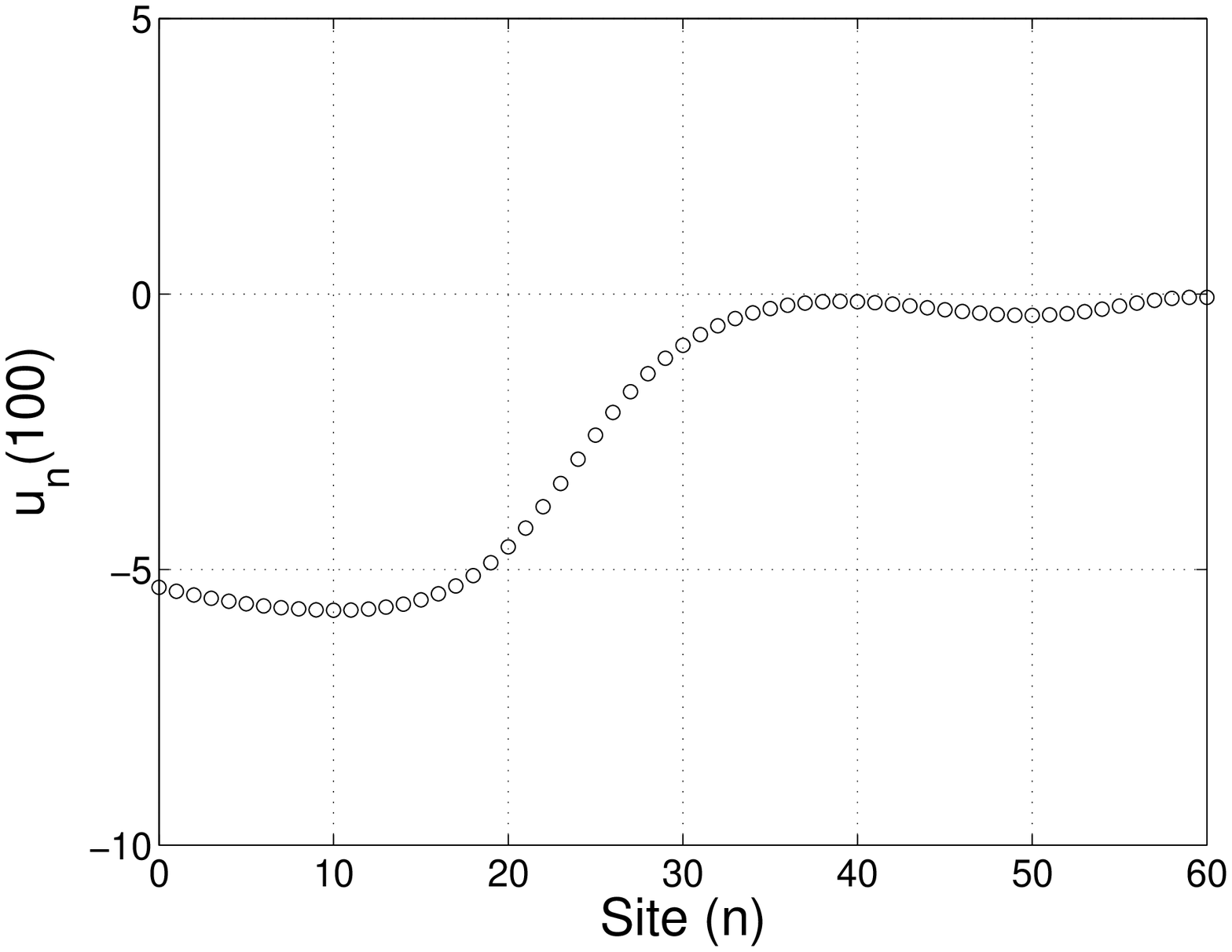}&\includegraphics[width=0.45\textwidth]{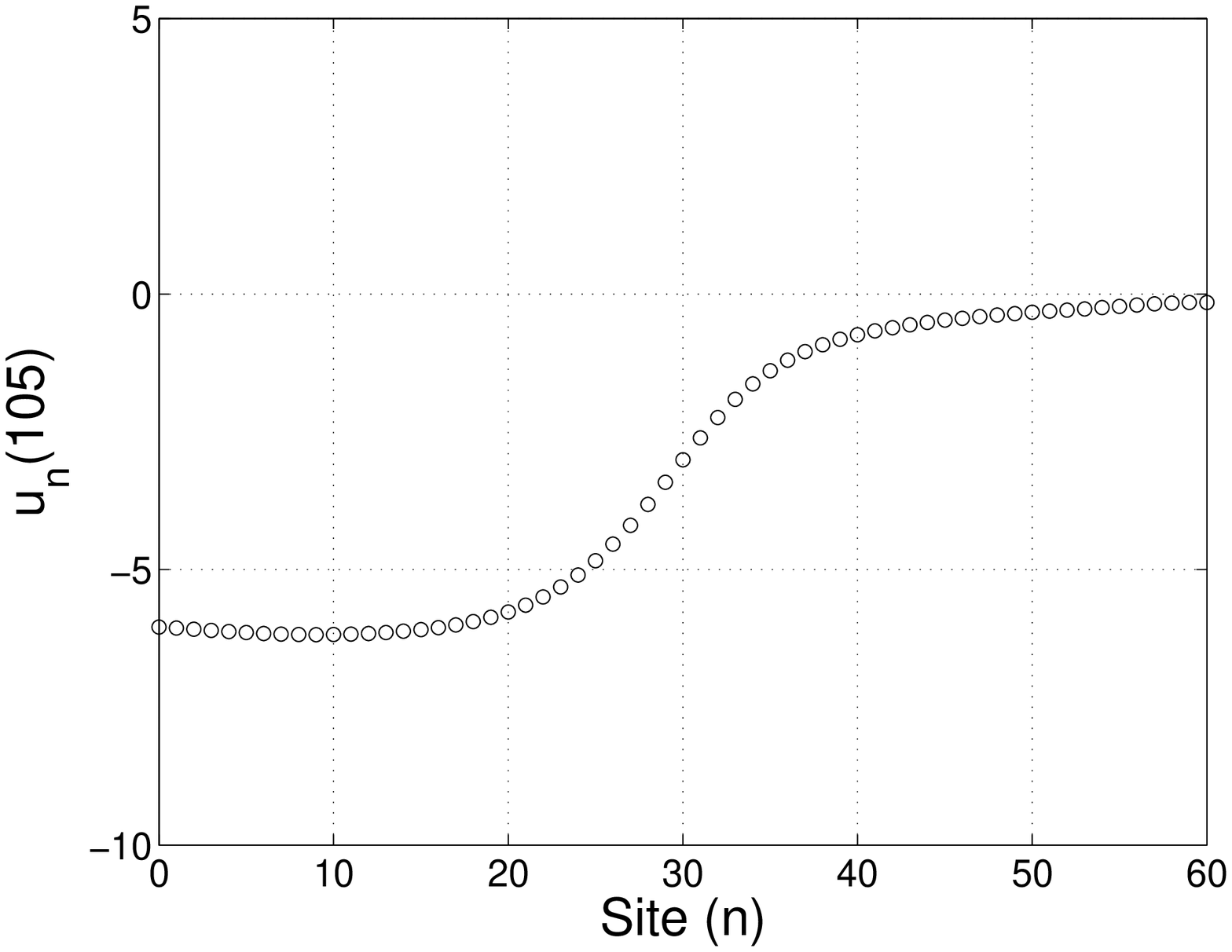}\\
\includegraphics[width=0.45\textwidth]{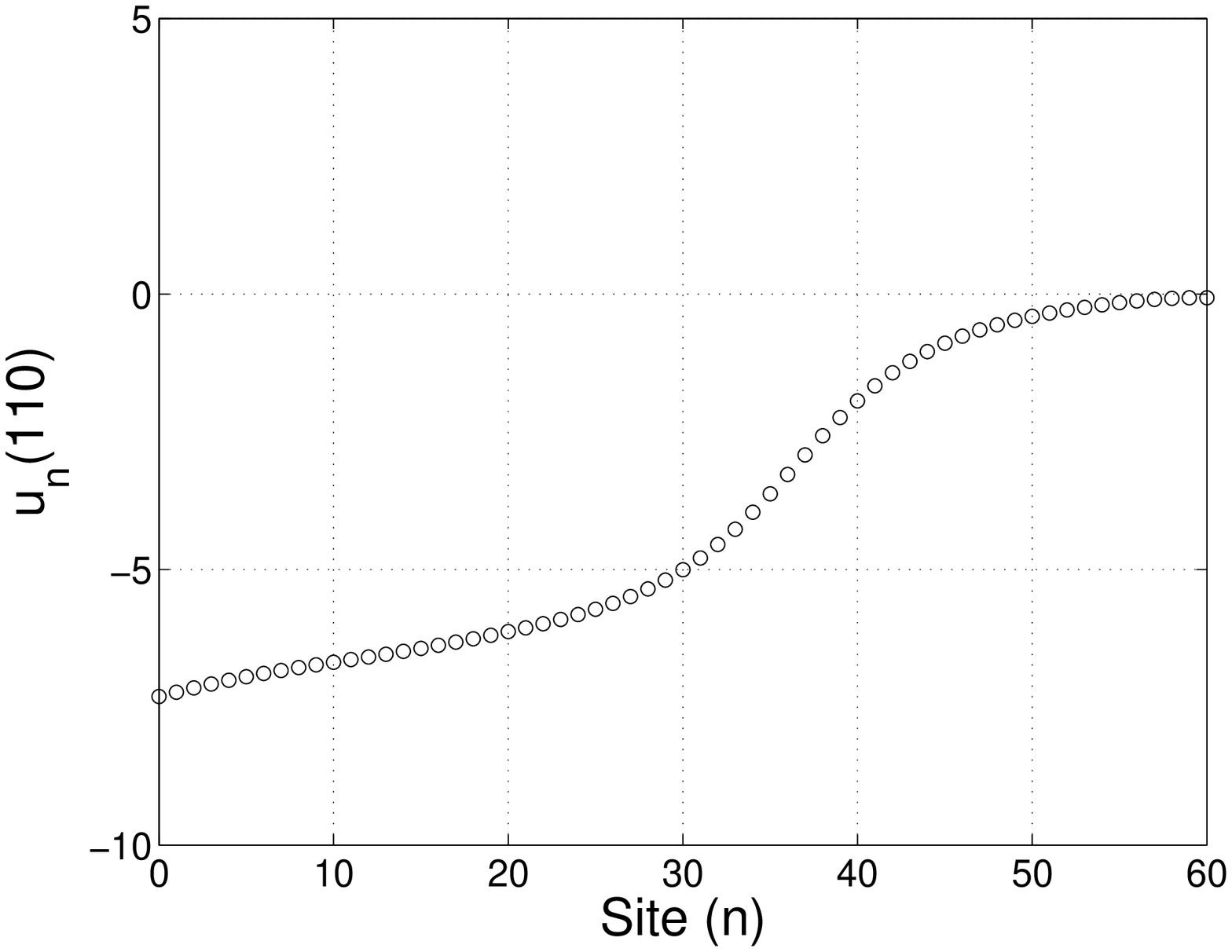}&\includegraphics[width=0.45\textwidth]{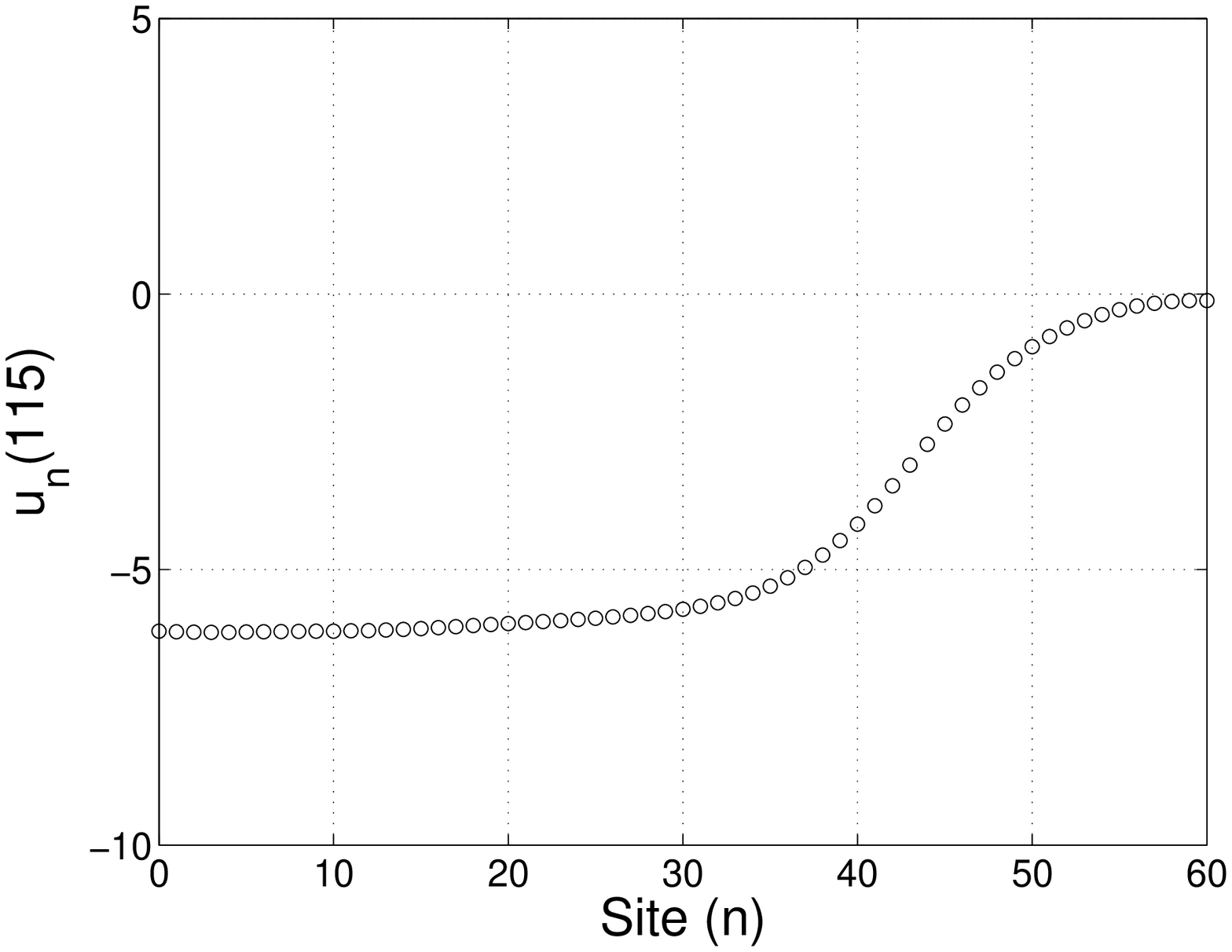}
\end{tabular}
\caption{Development of a kink solution of (\ref {Eqn:DiscreteMain20}) for discrete array of $60$ Josephson junctions, with $\gamma = \mu = 0$ and $c = 5$, driven at the end with by a frequency $\Omega = 0.9$ and an amplitude $A = 2$ slightly above the supratransmission threshold. The snapshots were taken at $6$ times equally spaced between $90$ and $115$, employing a step size of $0.2$ and an absorbing boundary in the last $10$ junctions.\label{Fig:Kinks}}
\end{figure}

\begin{figure}[tbc]
\begin{tabular}{cc}
\includegraphics[width=0.45\textwidth]{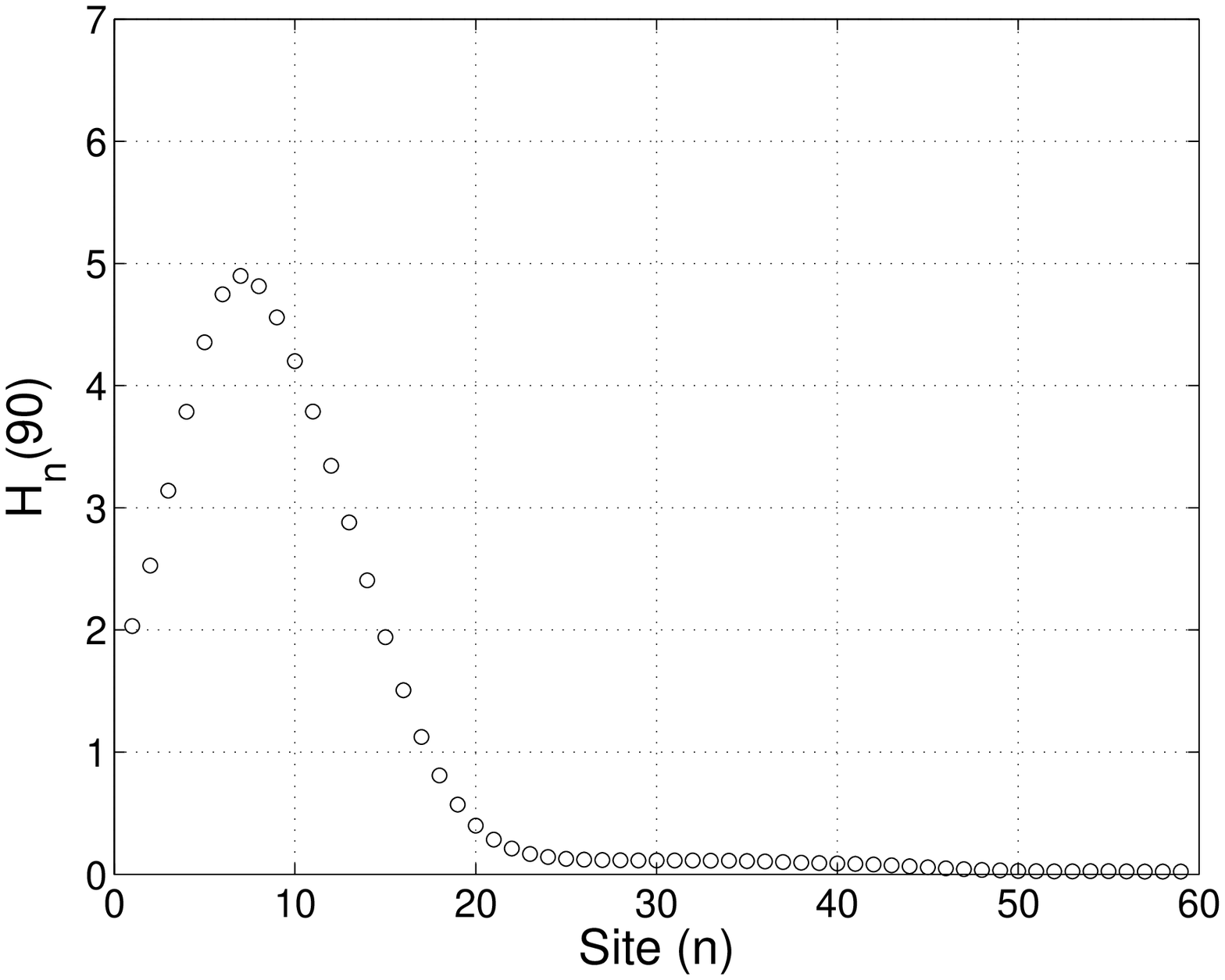}&\includegraphics[width=0.45\textwidth]{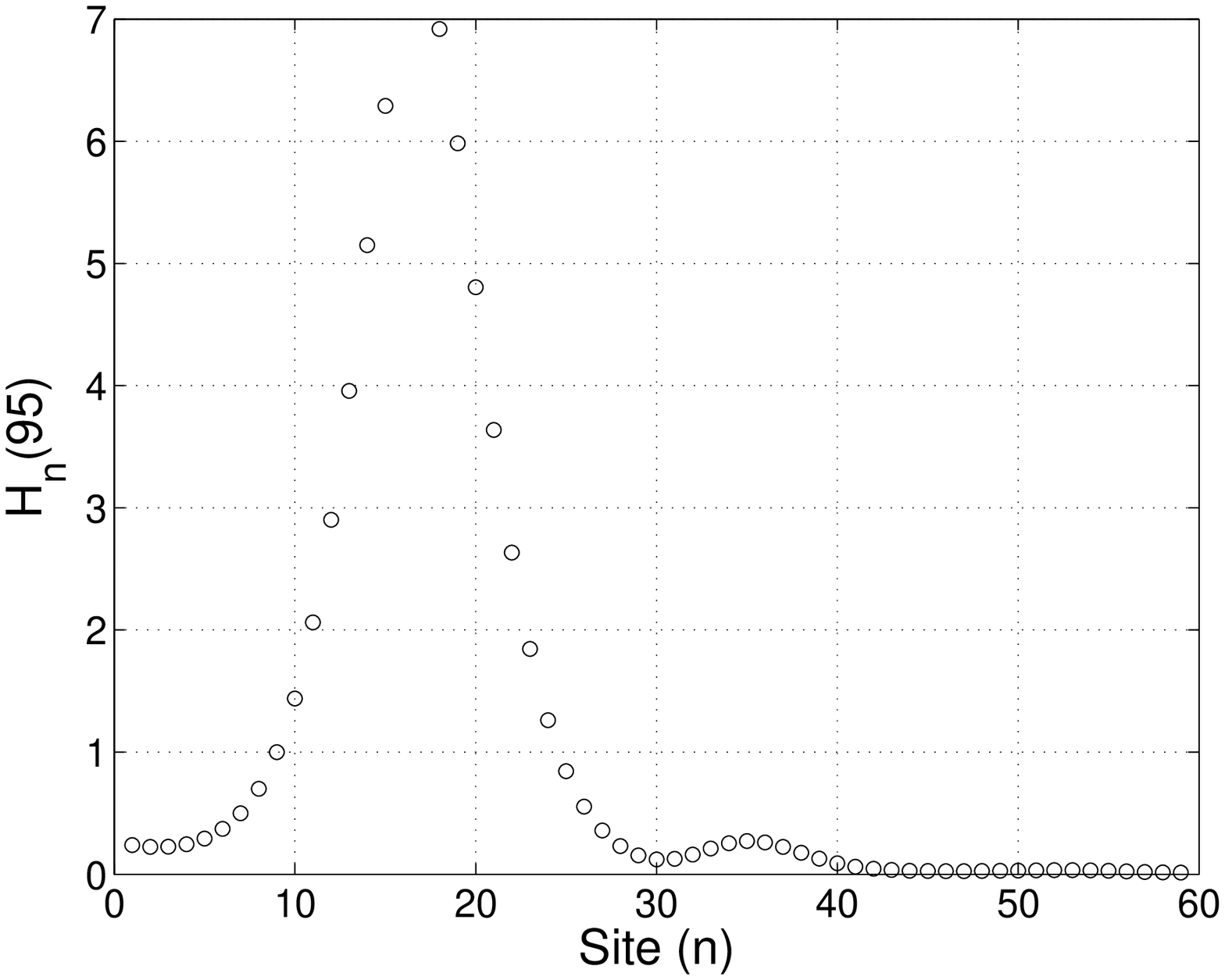}\\
\includegraphics[width=0.45\textwidth]{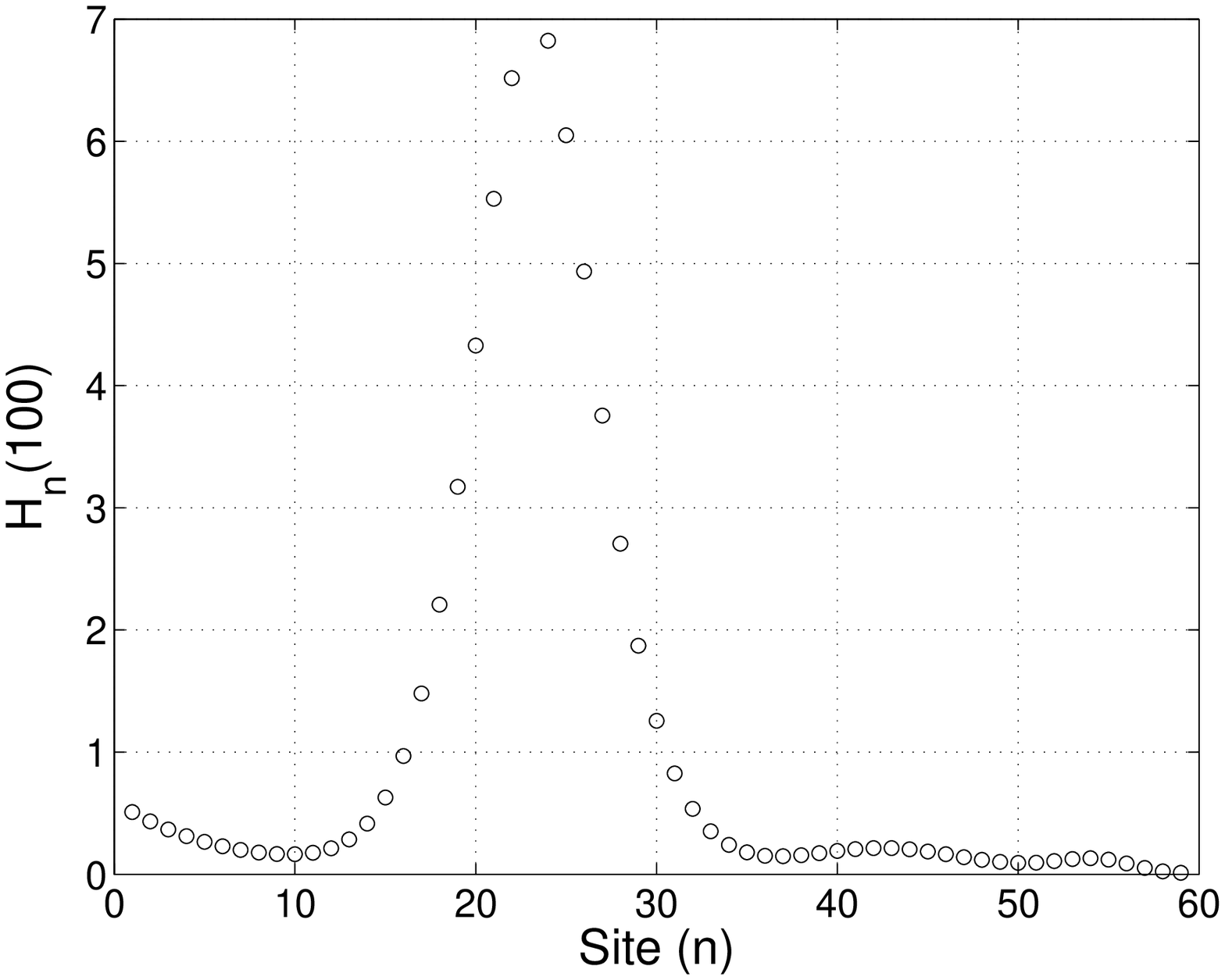}&\includegraphics[width=0.45\textwidth]{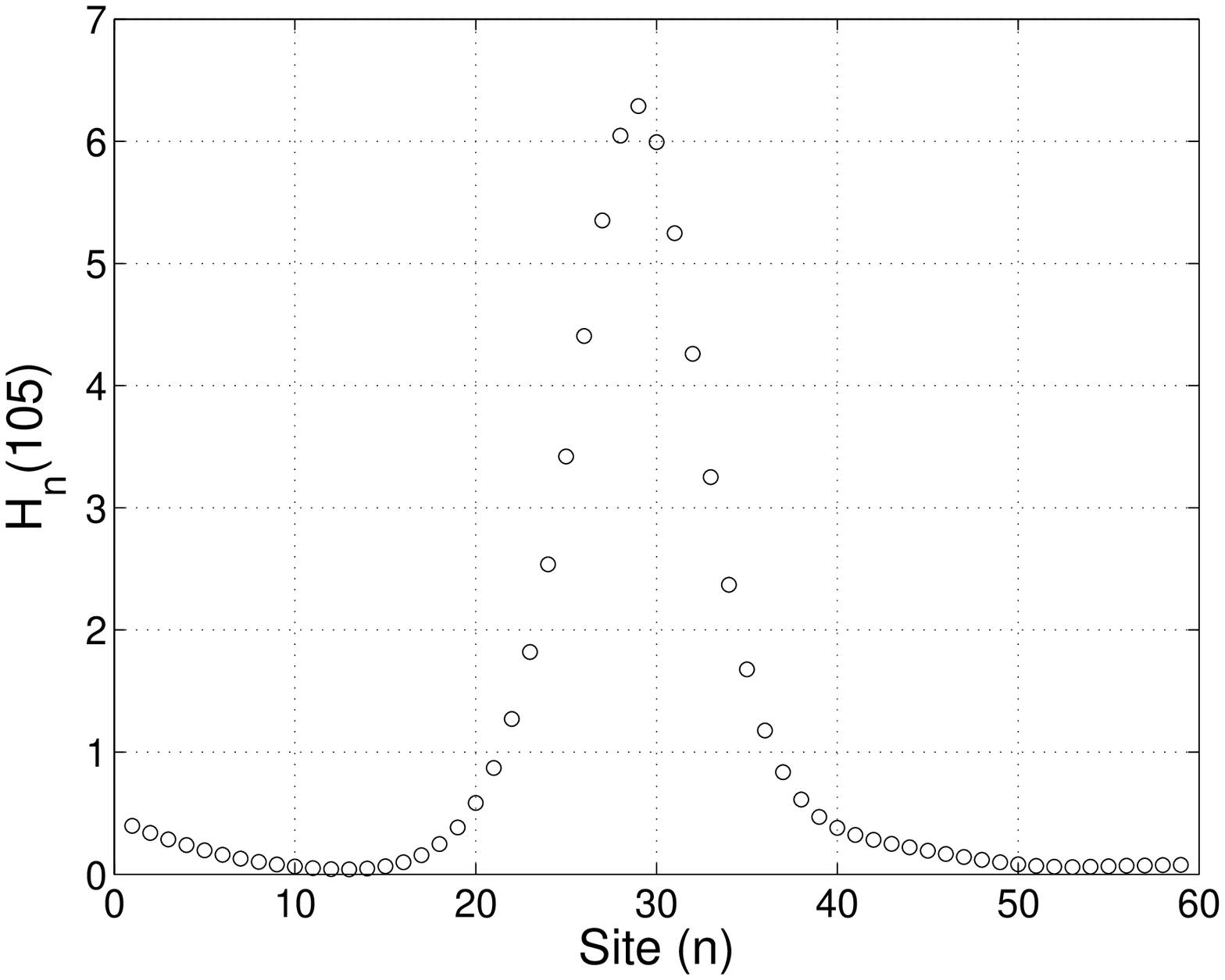}\\
\includegraphics[width=0.45\textwidth]{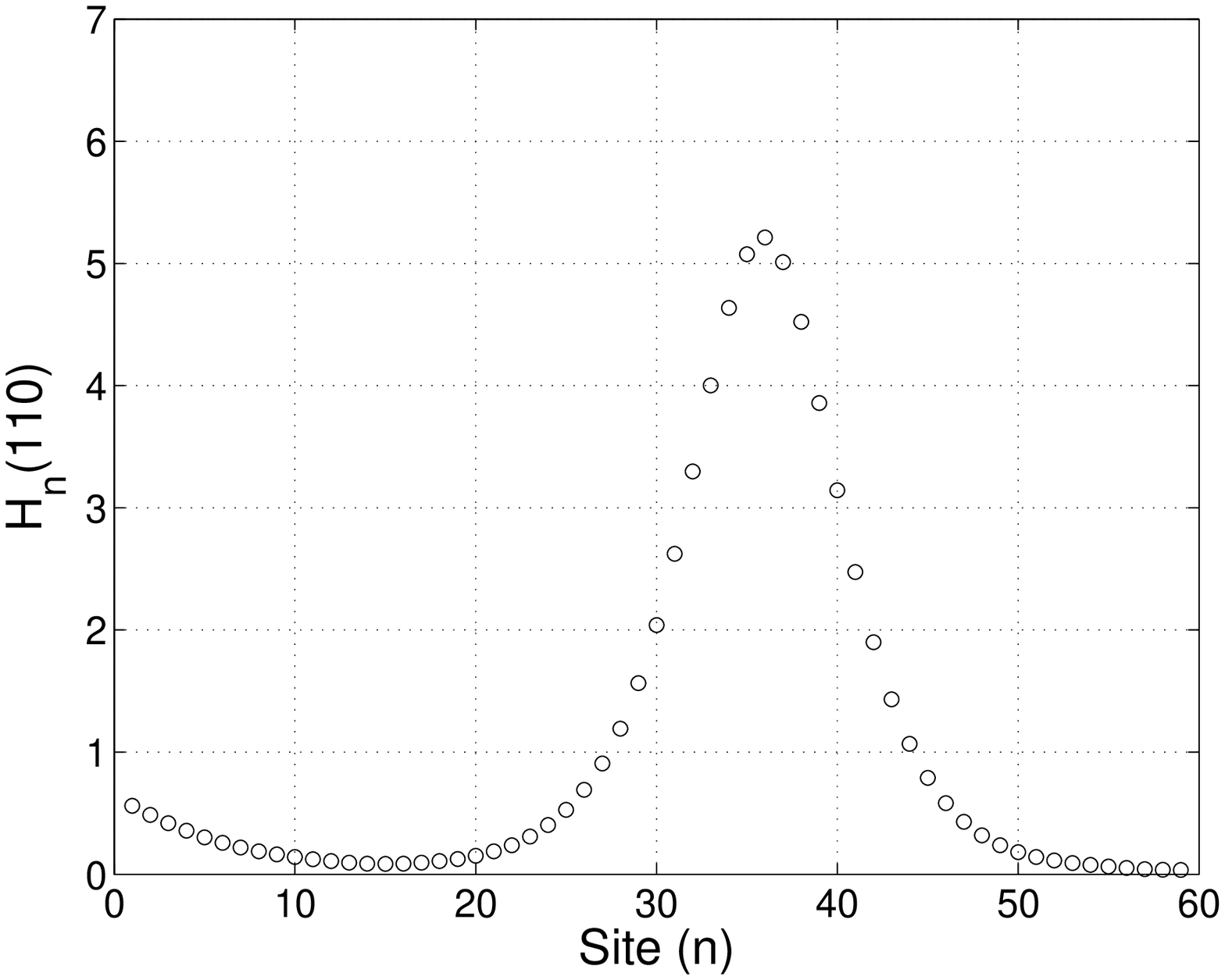}&\includegraphics[width=0.45\textwidth]{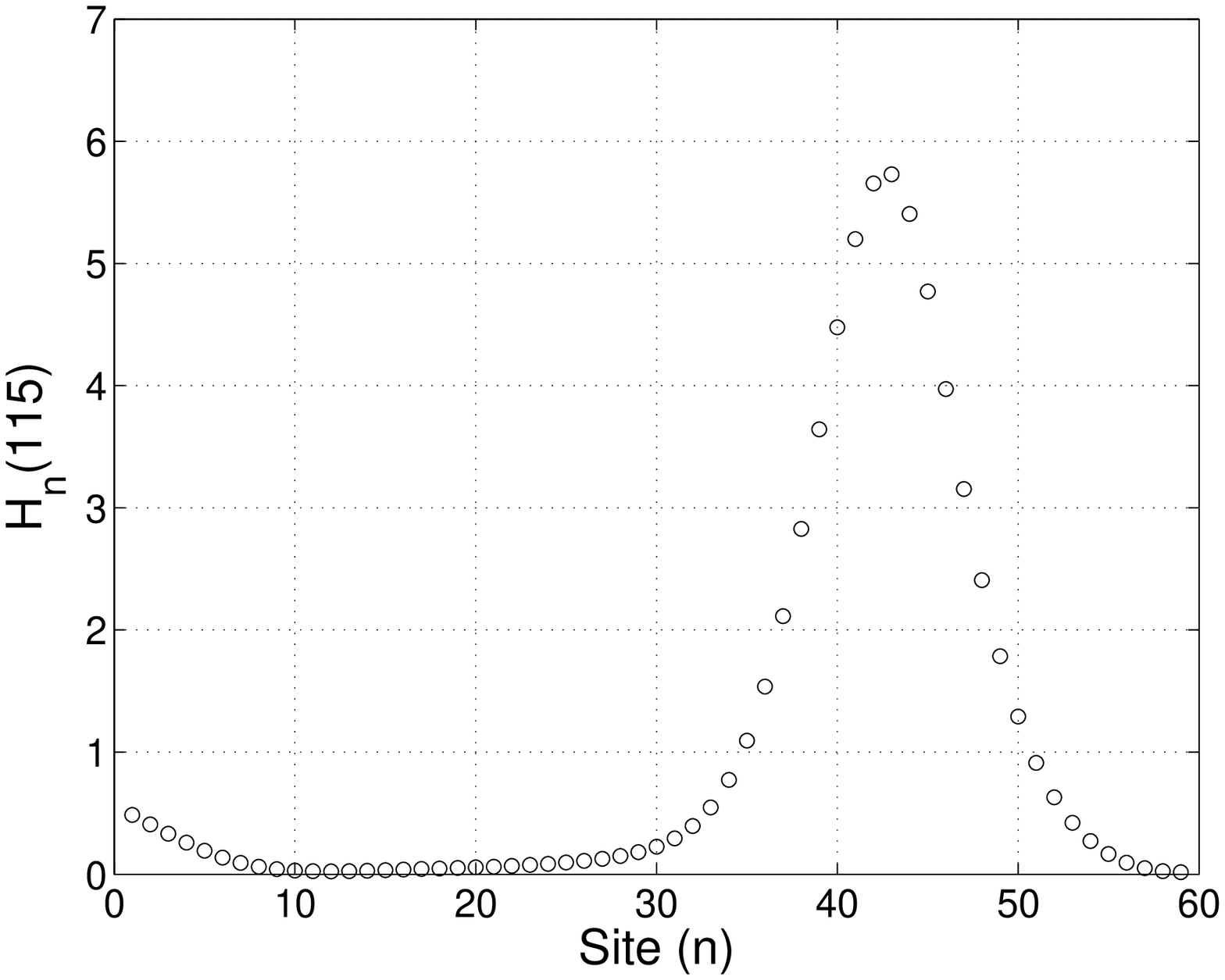}
\end{tabular}
\caption{Development of the local energies in a kink solution of (\ref {Eqn:DiscreteMain20}) for discrete array of $60$ Josephson junctions, with $\gamma = \mu = 0$ and $c = 5$, driven at the end with by a frequency $\Omega = 0.9$ and an amplitude $A = 2$ slightly above the supratransmission threshold. The snapshots were taken at $6$ times equally spaced between $90$ and $115$, employing a step size of $0.2$ and an absorbing boundary in the last $10$ junctions.\label{Fig:EnergyKinks}}
\end{figure}

\begin{figure}[tbc]
\begin{tabular}{ccc}
\includegraphics[width=0.45\textwidth]{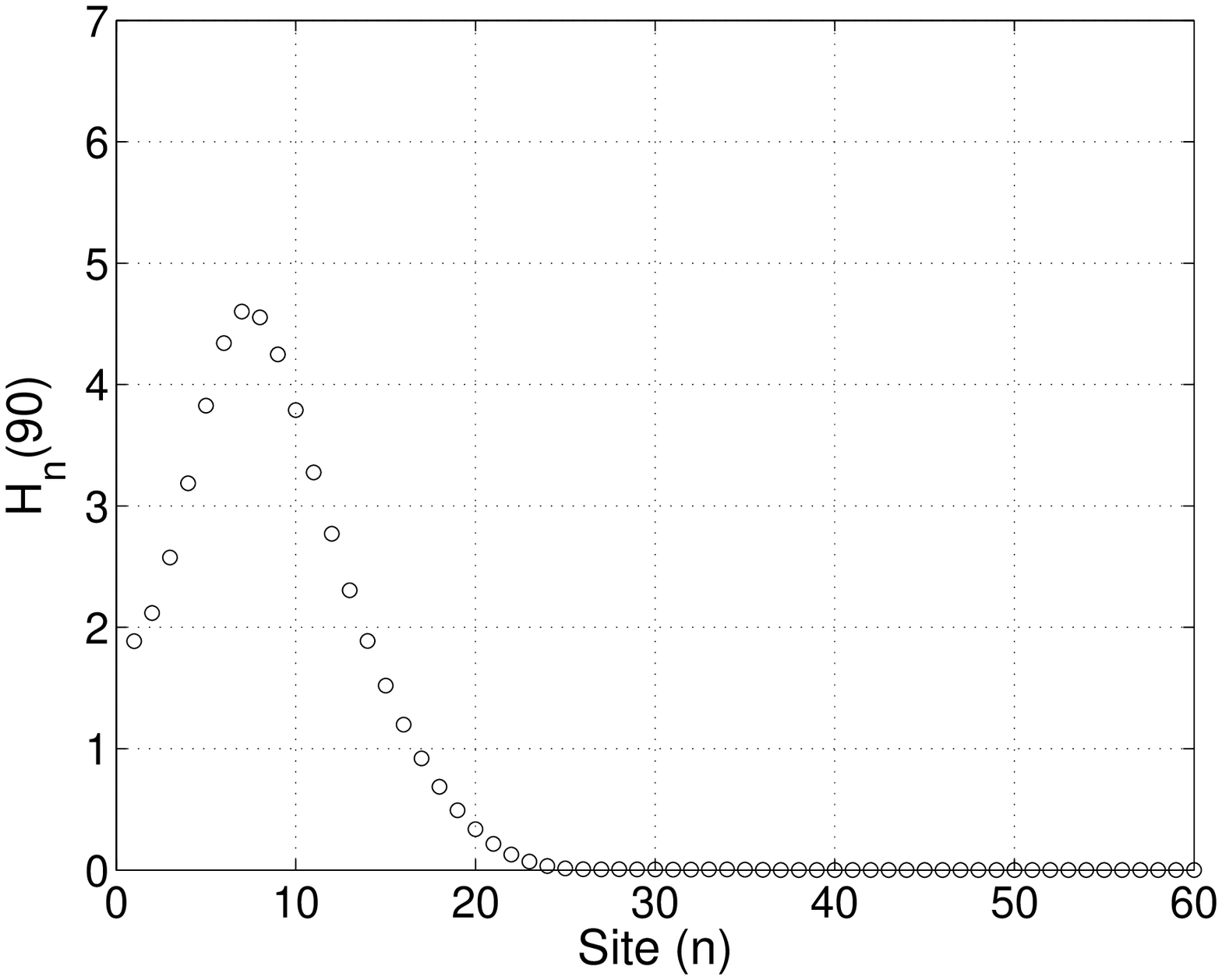}&\includegraphics[width=0.45\textwidth]{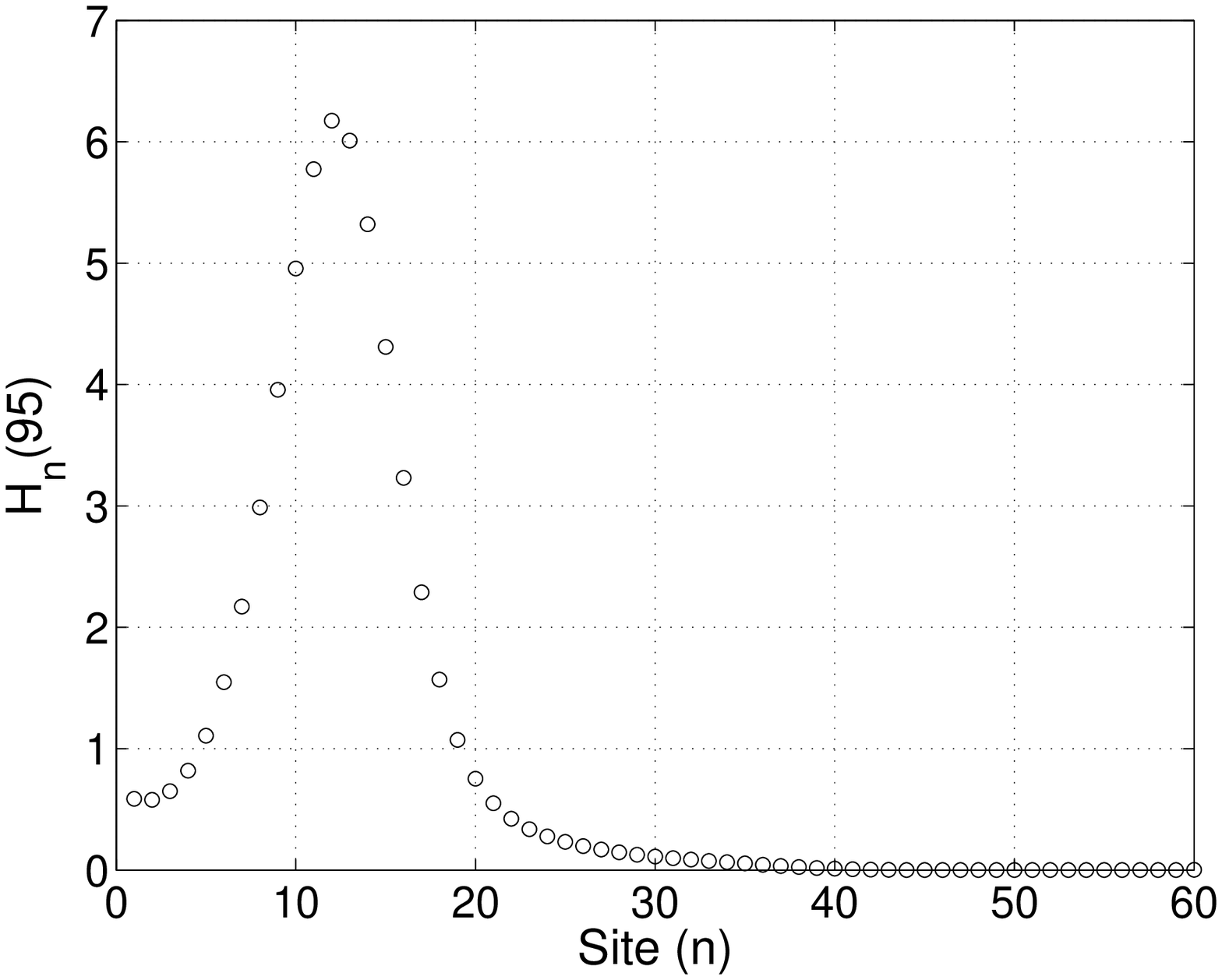}\\
\includegraphics[width=0.45\textwidth]{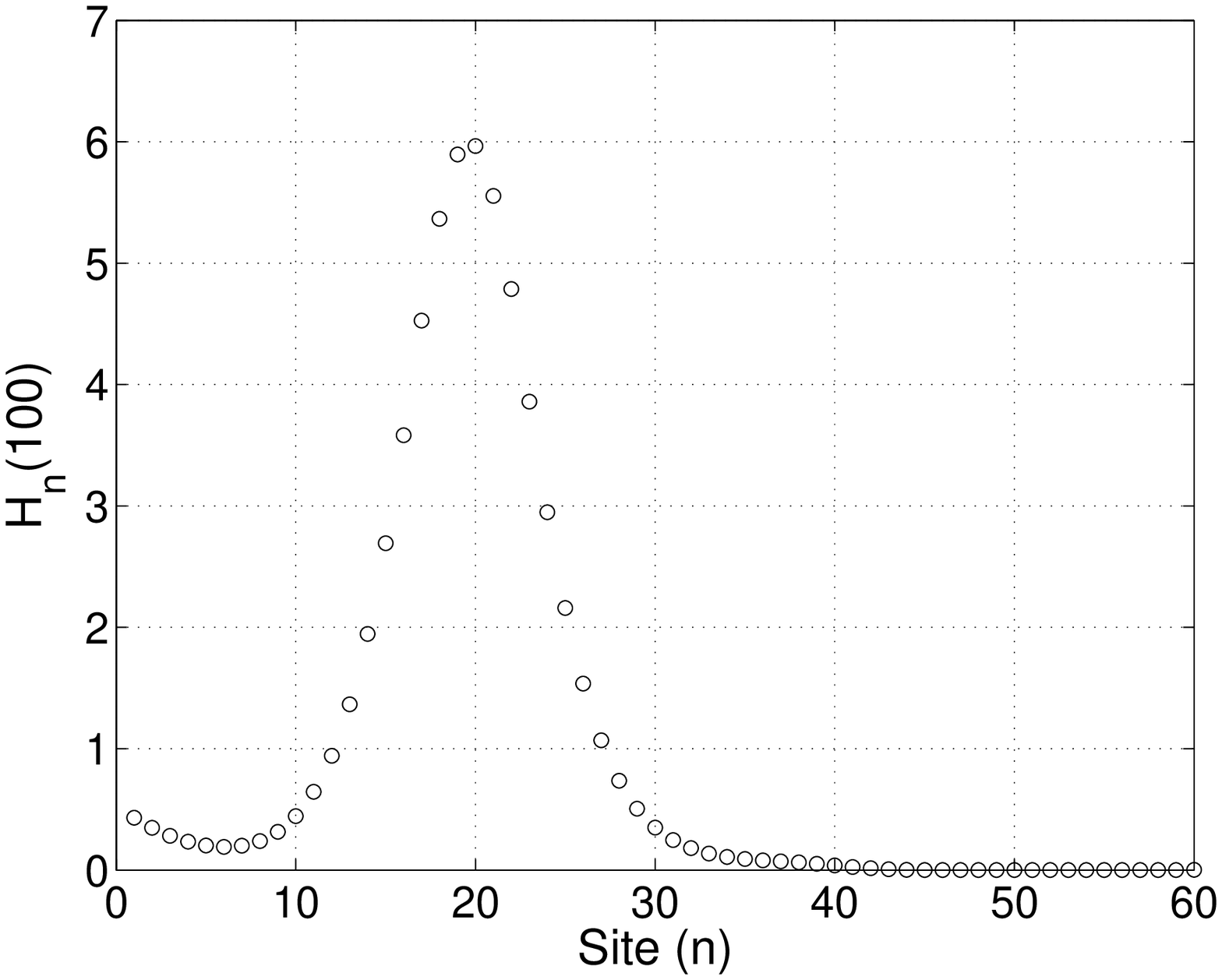}&\includegraphics[width=0.45\textwidth]{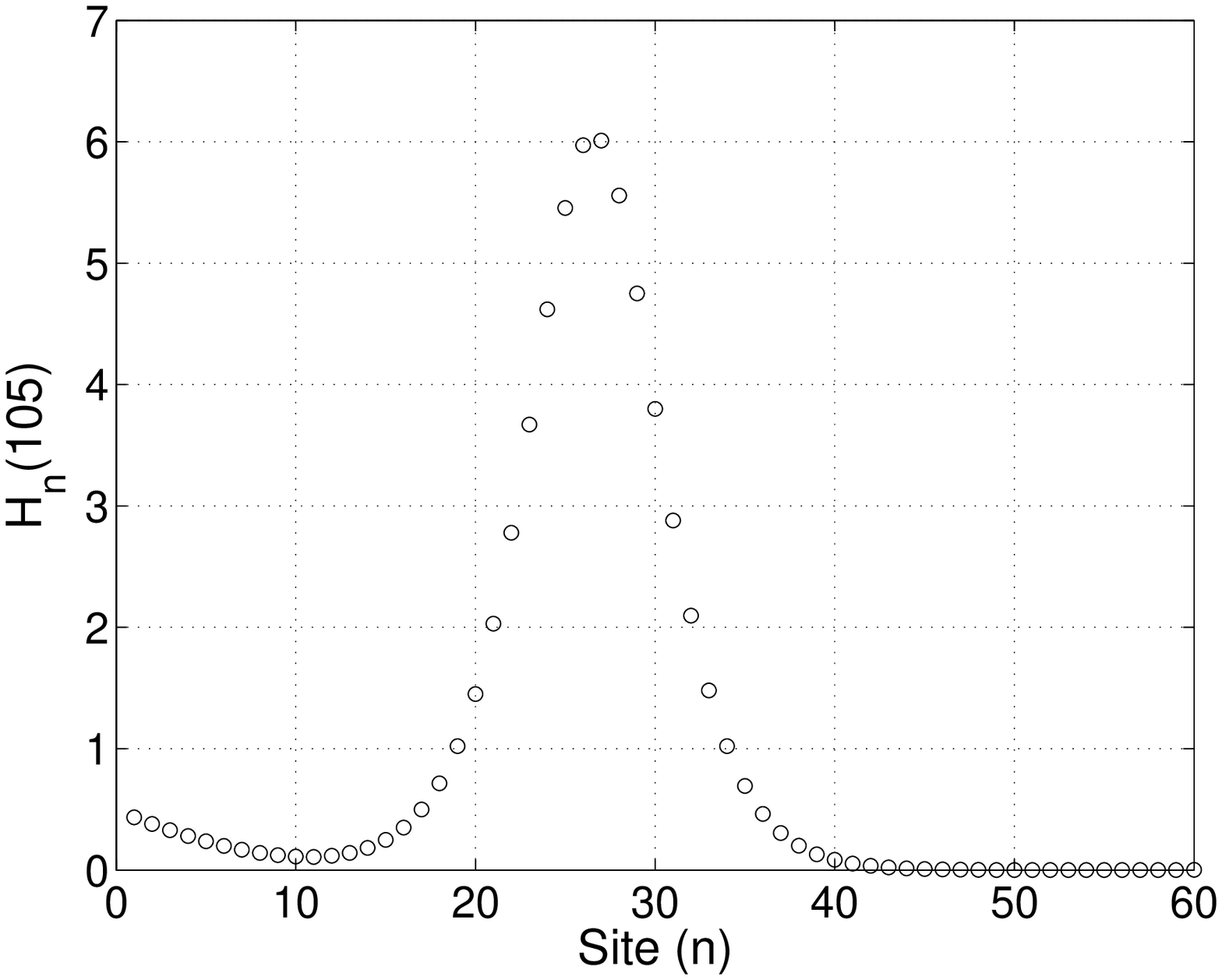}\\
\includegraphics[width=0.45\textwidth]{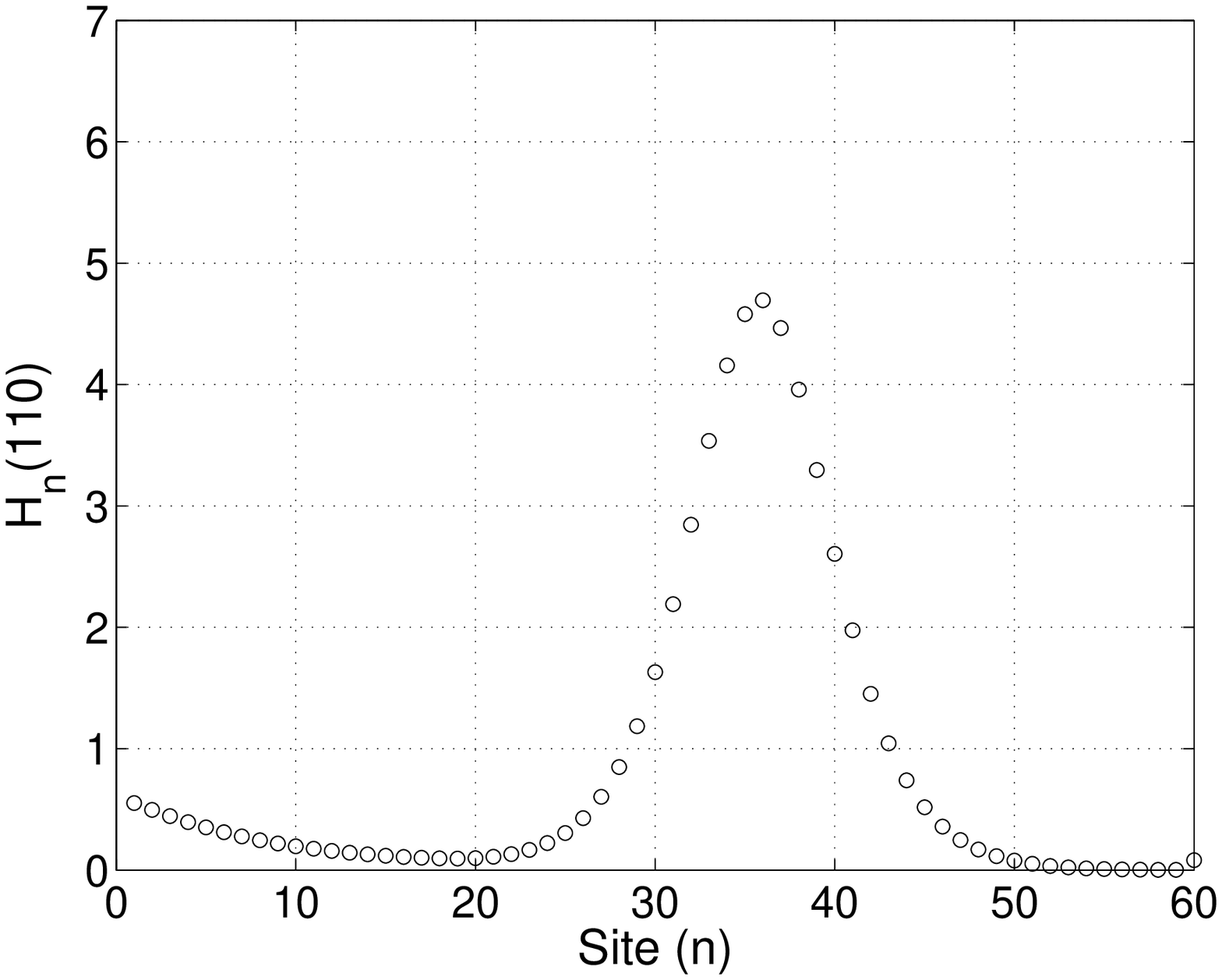}&\includegraphics[width=0.45\textwidth]{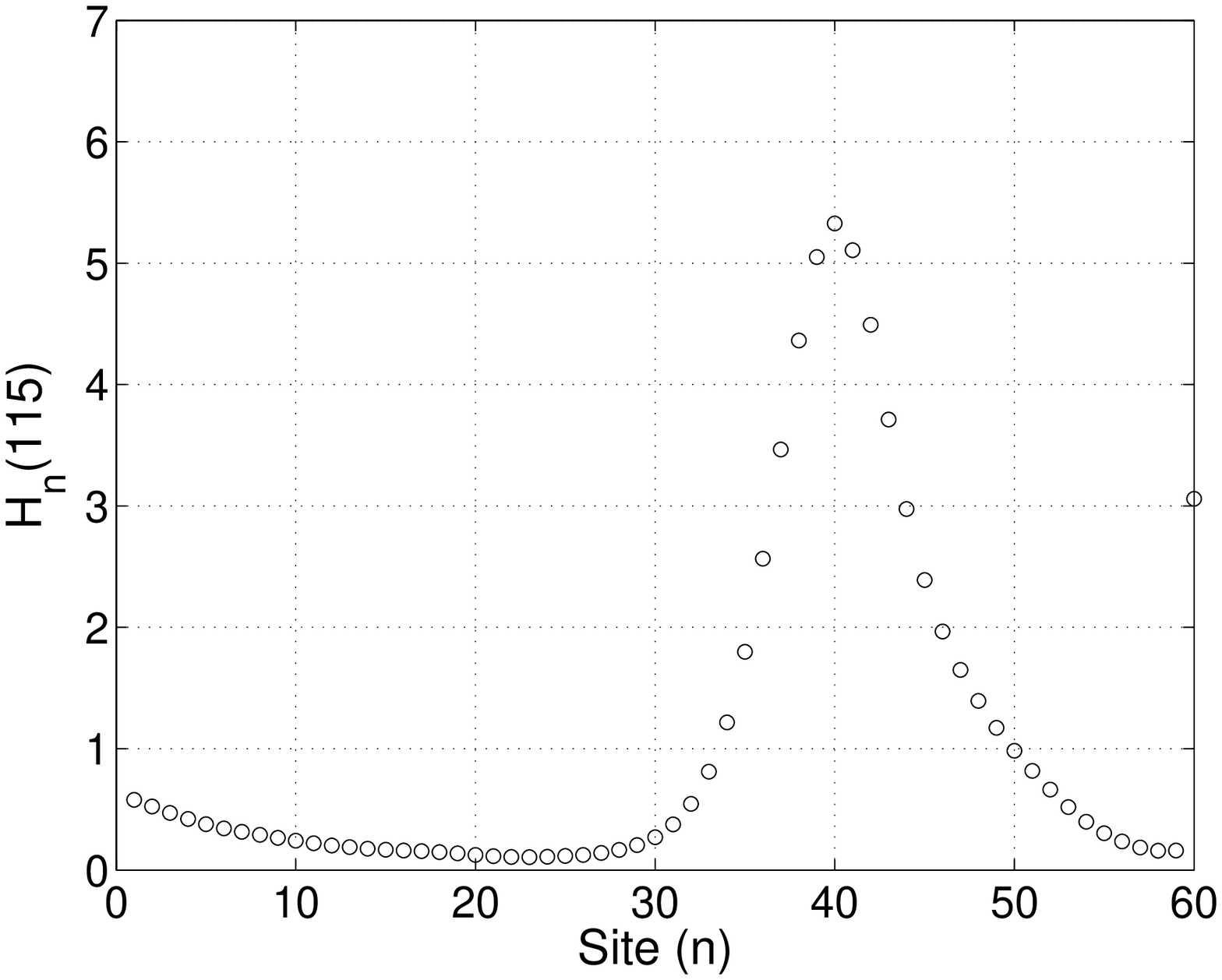}
\end{tabular}
\caption{Development of the local energies in a kink solution of (\ref {Eqn:DiscreteMain20}) for discrete array of $60$ Josephson junctions, with $\gamma = 0.001$, $\mu = 0$ and $c = 5$, driven at the end with by a frequency $\Omega = 0.9$ and an amplitude $A = 2$ slightly above the supratransmission threshold. The snapshots were taken at $6$ times equally spaced between $90$ and $115$, employing a step size of $0.2$ and an absorbing boundary in the last $10$ junctions.\label{Fig:DEnergyKinks}}
\end{figure}

\begin{figure}[tbc]
\centerline{\includegraphics[width=0.9\textwidth]{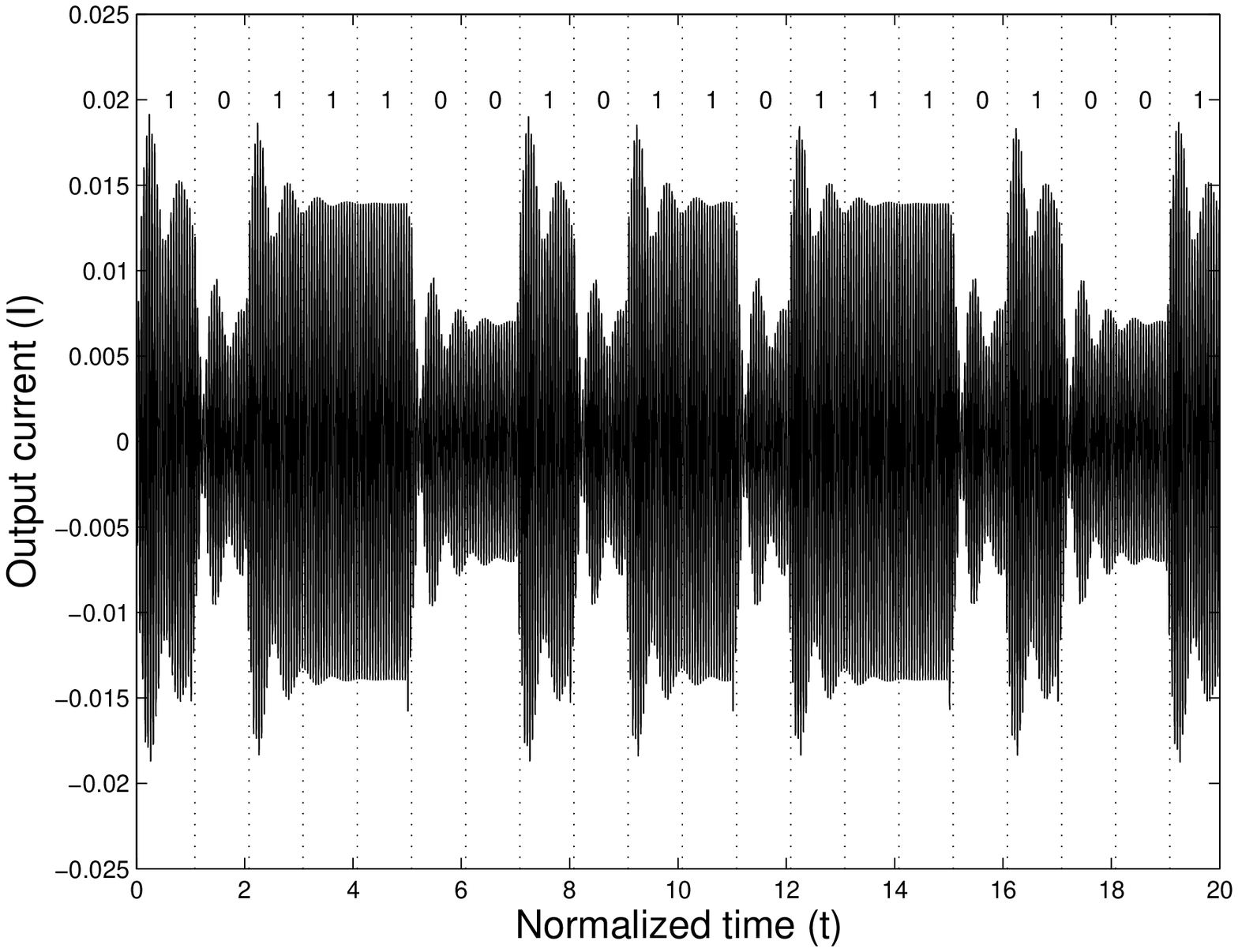}}
\caption{Graph of output current intensity vs. normalized time of system (\ref {Eqn:DiscreteMain20}) consisting of $8$ coupled junctions with $c = 2$, as a response of the input current function (\ref {Eqn:Input}) for the binary message `10111001011011101001'. The parameters $\gamma = 0.002$, $\Omega = 0.9$ and $R = 10$ are employed. \label{Fig:Paper0Fig2}}
\end{figure}

A finite time period of relatively large length is chosen, and a finite, undamped system consisting of $60$ Josephson junctions with no Josephson current, a coupling coefficient of $5$, and a driving frequency of $0.9$ is fixed, for which the nonlinear supratransmission threshold is predicted to be equal $1.945$, according to Equation (\ref {Eqn:PredictedBif}). Several driving amplitudes $A _0$ are employed, and the total energy of the system is computed for each amplitude. The results of driving the system with an amplitude function given by (\ref {Eqn:Amplitude}) with $\tau _1 = 10$ and $\tau _2 = 100$ are shown in Figure \ref {Fig:InfraGraph}, in which a drastic change in the behavior of the total energy with respect to $A _0$ is detected around the value $1.41$. This value is identified as the nonlinear infratransmission threshold of the system for $\Omega = 0.9$.

As in the case of nonlinear supratransmission, the critical amplitude value at which infratransmission occurs also delays with the value of $\gamma$. More precisely, the larger the value of the damping coefficient the larger the critical infratransmission value.

\section{Signal propagation\label{Sec4}}

\subsection{Character of solutions}

Consider a system of $60$ coupled oscillators described by (\ref {Eqn:DiscreteMain20}) with coupling coefficient equal to $5$, driven at a frequency of $0.9$ in the forbidden band gap, and an amplitude of $2$, just above the approximate supratransmission threshold. Figure \ref {Fig:Kinks} shows the time evolution of a kink moving away from the driving boundary. The exact location of the kink can be better determined by studying the time evolution of the local energies of the $60$ sites at the corresponding times.

With that purpose in mind, Figure \ref {Fig:EnergyKinks} presents the evolution of the local energies of the six snapshots in the previous figures. The location of the kinks is accurately identified, at each individual time, as the absolute maximum of the local energies. Moreover, a constant phase velocity of approximately 1.44 sites per unit of time is observed.

Finally, in Figure \ref {Fig:DEnergyKinks} we present a simulation of the evolution of the local energies at the same instants of time as in Figure \ref {Fig:EnergyKinks}. The same parametric values have been used in both figures, except that in this last graph we employed a coefficient of damping equal to $0.001$. In this situation, our numerical results (not presented here for the sake of briefness) show that the supratransmission threshold for this value of $\gamma$ is still smaller than the driving amplitude $2$. Indeed, our results show that there is propagation of energy in the form of localized modes; however, dissipation of energy is present too, as evidenced by the fact that the local energy density of localized solutions is slightly smaller when $\gamma$ is not zero. 

In both of the cases presented here, the group velocity of the kink produced by the driving boundary is approximately the same: 1.44. However, the signal of the dissipative case seems to be a little delayed with respect to the case then damping is not present. This observation is in perfect agreement with the fact that the driving amplitude $A$ is slowly and linearly increased from $0$ to $2$, in order to avoid the creation of shock waves, so that the critical amplitude for the case $\gamma = 0$ is reached before the critical value for a nonzero $\gamma$.

\subsection{Simulation}

Let $\Omega$ be a frequency in the forbidden band gap of our problem. Assume that $B _i$ and $B _s$ are nonnegative values that are just a bit smaller than the infratransmission and supratransmission thresholds, respectively, both associated to the frequency $\Omega$. We define the \emph {seed} of the system as the function
\begin{equation}
I _s (t) = \frac {1} {2} \left[ (B _s - B _i) \sin (\Omega t) + B _i + B _s \right], \label{Eqn:Seed}
\end{equation}
for every $t > 0$. 

Next, we define the \emph {period of signal generation} $P$ as an integer multiple of the driving period. In our case, $P$ will be equal to $20$ times the period of driving. 

Let $\alpha$ be a positive number with the property that $I _s (t) + \alpha$ overcomes the nonlinear supratransmission threshold for some values of $t$. With these conventions in mind, a single binary message $(b _1 , b _2 , \dots , b _l)$ consisting of $l$ binary bits will be transmitted into system (\ref {Eqn:DiscreteMain20}). In general, for every $m = 1 , 2 , \dots , l$, we define the signal function $S _m$ by 
\begin{equation}
S _m (t) = \left\{\begin{array}{ll}
\alpha b _m \sin (\Omega t), & {\rm if} \ m P < t < (m + 1) P, \\
0, & {\rm otherwise}.
\end{array}\right.
\end{equation}
The input intensity function will be defined then by
\begin{equation}
\phi (t) = I _s (t) + \sum _{m = 1} ^l S _m (t). \label{Eqn:Input}
\end{equation}

Consider a discrete Josephson junction arrays consisting of $8$ junctions with coupling coefficient equal to $2$. A driving frequency $\Omega = 0.9$ will be employed, for which the values $B _i = 0.23$, $B _s = 0.41$, $\alpha = 0.05$, $\gamma = 0.002$ and $R = 10$ will be used. The binary sequence `10111001011011101001' is to be transmitted into system (\ref {Eqn:DiscreteMain20}) by modulating the input current intensity and read by the corresponding output current. The resulting outcome of our simulations is summarized in Figure \ref {Fig:Paper0Fig2} for a normalized time.

Nonzero bits are clearly identified with output intensities of absolute values higher than the corresponding intensities associated with bits equal to zero. More accurately, nonzero bits are completely characterized by the fact that, at some time in the corresponding period of reception, the value of the intensity of the output signal is higher than a cutoff limit (in this case equal to $0.01$).

Several other experiments have been carried out for different values of $\gamma$. Here, it is interesting to notice that our experiments have produced results which are qualitatively in agreement with that presented in Figure \ref {Fig:Paper0Fig2}. Particularly, it is important to mention that there is a well-defined difference between the bits $0$ and $1$. The only difference is the amplitude of the oscillations of the output current, which decreases as $\gamma$ is increased, as expected.

\section{Conclusions and discussion}

In this letter, we have proved, using numerical computations, that it is possible to transmit binary information in discrete Josephson junction arrays using the processes of nonlinear supratransmission and infratransmission. In the absence of dissipative effects, our model (which is based on the modulation of amplitudes of source signals with constant frequency) has shown to be highly reliable for sufficiently long periods of single-bit generation. Moreover, our computations show that the general picture does not change much when weak damping is present; indeed, the only difference obtained between the conservative and the dissipative cases lies in the amplitude of the oscillations of the output current intensity. In practice, the knowledge of the speed at which localized solutions move through a discrete array consisting of Josephson junctions in parallel and the relation between driving amplitude, supratransmission and infratransmission thresholds and damping, may be fruitful in the accurate design of binary signal transmitters, as well as in digital amplifiers and signal detectors.

It is worth noticing the similarities and differences of our results with respect to the corresponding Dirichlet boundary-value problem. The results derived in this paper in the supratransmission analysis for the external damping are qualitatively similar to those presented in \cite{Macias-Supra}. On the other hand, the fundamental structures for transmission of wave signals in the Dirichlet scenario involved the propagation of moving breathers \cite{Macias-Signals,Macias-Signals2}; meanwhile, kinks and anti-kinks have been generated in the Neumann boundary-value problem.

\subsection {Acknowledgments}

One of us (JEMD) wishes to express his gratitude to Dr. \'{A}lva\-rez Rodr\'{\i}guez, dean of the Faculty of Sciences of the Universidad Aut\'{o}noma de Aguascalientes, and to Dr. Ave\-lar Gonz\'{a}lez, head of the Office for Research and Graduate Studies of the same university, for providing him with the means to produce this paper. He also wishes to thank the anonymous reviewer for all the invaluable comments that led to improve the quality of this work. The present article represents a set of partial results under project PIM07-2 at this university.

\end{document}